\documentstyle[floats,prd,aps,epsf,eqsecnum,12pt]{revtex}

\makeatletter
\newbox\tempboxa
\newdimen\captionboxsubcount
\def\capsize#1{\captionboxsubcount=#1pt}
\newdimen\captionboxsub
\captionboxsub=\hsize \advance\captionboxsub by -\captionboxsubcount
\advance\captionboxsub by -\captionboxsubcount
\long
\def\@makecaption#1#2{
\setbox\@tempboxa\hbox{#1 #2}
\ifdim \wd\@tempboxa >\captionboxsub
\rightskip=\captionboxsubcount \leftskip=\captionboxsubcount #1 #2
\else \hbox to\hsize{\hfil\box\@tempboxa\hfil}
\fi}
\makeatother
\capsize{30}

\begin{document}

\begin{titlepage}
\begin{flushright}
\begin{minipage}{5cm}
\begin{flushleft}
\small
\baselineskip = 13pt
YCTP-P7-99\\
hep-th/9903416 \\
\end{flushleft}
\end{minipage}
\end{flushright}
\begin{center}
\Large\bf
Cosmological Neutrino Condensates
\end{center}
\vfil

\begin{center}
\large
D.G. {\sc Caldi}\footnote{Electronic address: {\tt
caldi@caldi.physics.buffalo.edu}} \\ {\it \qquad Department of
Physics, State University of New York at Buffalo, Buffalo, NY
14260} \\

Alan {\sc Chodos}\footnote{ Electronic address : {\tt
chodos@hepvms.physics.yale.edu}}\\ { \it \qquad Department of
Physics, Yale University, New Haven, CT 06520-8120, USA}\\

\vskip .5cm
 {\sc }

\qquad

\end{center}
\vfill
\begin{center}
\bf
Abstract
\end{center}
\begin{abstract} We investigate the possibility that neutrinos form
superfluid-type condensates in background cosmological densities. Such
 condensates could give rise to small neutrino masses and
splittings, as well as an important contribution, perhaps, to the
cosmological constant. We discuss various channels in the context
of the standard model. Many of these do not support a condensate,
but some mixed-flavor channels do. We also suggest a new
interaction, acting only among neutrinos, that could induce a
neutrino Majorana mass of order $1$ eV.
\baselineskip = 17pt

\end{abstract}
\begin{flushleft}

PACS numbers:
\end{flushleft}
\vfill
\end{titlepage}

For various reasons, but especially the recent evidence for
neutrino oscillations and hence neutrino mass [1], considerable
effort has been devoted of late to neutrinos and their properties.
Furthermore, another more speculative consideration has piqued
interest in small neutrino masses: ~a number of cosmological
observations has led to renewed interest in the long-discarded
notion of a cosmological constant [2]. The preferred value of this
constant leads to a mass scale that is quite similar to the mass
for neutrinos inferred from the atmospheric and solar neutrino
observations, in the range of tenths of an eV to a few eV [1]. It
seems to us that a natural framework in which to relate these
seemingly disparate phenomena would be the formation of some type
of neutrino condensate, the energy of which might provide the
cosmological constant, while the neutrino masses could emerge from
an expansion about the symmetry-breaking vacuum.

Appropriate tools to investigate the possibility of neutrino pair
condensation of the superconductor type, have been recently
employed to explore the possibility of $qq$ condensates in hadronic
media at sufficiently high densities. This has been discussed in
the context of a four-fermi approximation to $QCD$, whether induced
by instanton effects [3]or by one-gluon exchange [4]. Similar
effects have also been studied in the large-$N$ approximation in a
$2$-dimensional model related to the Gross-Neveu model [5].

For our purposes, we look for pairing phenomena in the electroweak
theory. Specifically, we suggest that condensates involving
neutrinos may occur in cosmological situations where the relevant
chemical potentials are non-zero. Of course, the chemical
potentials of the various neutrinos in the universe are presumably
quite tiny [6], reflecting as they do the difference in density
between neutrinos and anti-neutrinos of a given species. We shall
return to the estimation of the magnitude of the effects we discuss
below; first we offer a purely theoretical discussion of the
following problem: to identify, if any, the attractive channels in
the electroweak theory that might permit neutrinos to condense. It
should be noted that the earliest speculation on this subject of
which we are aware [7], took place before the emergence of the
standard model and the discovery of neutral currents, and the
authors simply assumed the existence of an attractive channel.

The hallmark of this type of pairing is that, once a Fermi surface
forms, any attractive channel, no matter how weak, can produce a
condensate; this is in contrast to the more familiar situation in
hadronic physics of a chiral condensate, which typically requires
the coupling to exceed a certain threshold value.

The reason for this behavior of the pairing condensate can be
understood in a variety of ways.  If one looks at the gap equation
in a mean-field approximation one sees that, as the gap tends to
zero, a singularity develops at the Fermi surface; hence a non-zero
gap is necessary to prevent the formation of this singularity.

Alternatively, in a renormalization-group approach [8] one sees
that the renormalized coupling tends to infinity as one integrates
out all modes above and below the Fermi surface; thus even an
apparently weak coupling grows without bound and allows the
condensate to form.

Our investigation will be limited to the leptonic sector of the
electroweak theory.  We shall use the effective four-Fermi
low-energy description of that theory and we shall perform all our
computations within the one loop (mean-field) approximation.
Corrections to this picture from the inclusion of effects not
captured by the low-energy effective theory, or from fluctuations
about the mean-field approximation, must await further
investigations. We shall discuss possible consequences of
extensions of the dynamics beyond the standard model below.

To set the stage, let us consider an action with a generic four
Fermi coupling:

$$
~~~~~~~~~~S = {1 \over 2} [\psi_{\alpha}^{\dagger}
A_{\alpha\beta}\psi_{\beta}
-
\psi_{\alpha}A^T_{\alpha\beta}\psi_{\beta}^{\dagger}] +
{\cal M}_{\alpha\beta\gamma\delta}
\psi_{\alpha}^{\dagger}\psi_{\beta}\psi_{\gamma}^{\dagger}\psi_{\delta} ~.
$$

\bigskip\noindent
Here the index $\alpha$ summarizes all the attributes of $\psi$:
~space-time dependence, gauge transformation properties, flavor,
spin and whatever else.

Because we are interested in pairings of the form $\langle
\psi\psi\rangle$ and $\langle
\psi^{\dagger}\psi^{\dagger}\rangle$, we assume that ${\cal M}$ has a
Fierz-Bogoliubov (FB) decomposition:

$$
~~~~~~~~~~~~{\cal M}_{\alpha\beta\gamma\delta} = \sum_{\lambda}
\eta_{\lambda} Q_{\alpha\gamma}^{(\lambda)}
Q_{\beta\delta}^{*(\lambda)}
.
$$

\bigskip\noindent
Here $\eta_{\lambda} = \pm 1$, and $Q_{\alpha\gamma}^{(\lambda)} =
- Q_{\gamma\alpha}^{(\lambda)}$ because of the Fermi statistics of
$\psi$. Then we define auxiliary fields $B^{(\lambda)}$, and add to
${\cal L}$ the term

$$
~~~~~~~~~- \sum_{\lambda} \eta_{\lambda} (B^{(\lambda)^{\dagger}} -
Q_{\alpha\gamma}^{(\lambda)}\psi_{\alpha}^{\dagger}
\psi_{\gamma}^{\dagger}) (B^{(\lambda)} + Q_{\beta\delta}
^{*(\lambda)} \psi_{\beta} \psi_{\delta}) ~.
$$

\bigskip\noindent
The original Lagrangian is recovered upon path-integrating on
$B^{(\lambda)}$ and $B^{(\lambda)\dagger}$. With the addition of
this extra piece, the terms in ${\cal L}$ that are quartic in
$\psi$ and $\psi^{\dagger}$ cancel, leaving us with

\vfill\eject
\begin{eqnarray*}
S &=& {1 \over 2} [\psi^{\dagger} A \psi - \psi A^T \psi^{\dagger}]
- \sum_{\lambda} \eta_{\lambda} B^{\dagger(\lambda)} B^{(\lambda)}
\\
 &+& \psi^{\dagger} {\cal B} \psi^{\dagger} + \psi
{\cal B}^{\dagger} \psi \ ,  \\
 {\cal B}_{\alpha\gamma} &=& \sum_{\lambda}
\eta_{\lambda} B^{(\lambda)} Q_{\alpha\gamma}^{(\lambda)} \ ,  \\
{\cal B}_{\beta\delta}^{\dagger} &=& - \sum_{\lambda}
\eta_{\lambda} B^{\dagger(\lambda)} Q_{\beta\delta}^{*(\lambda)} ~.
\end{eqnarray*}

\bigskip\noindent
We obtain an effective action $\Gamma_{eff}[B^{(\lambda)},
B^{(\lambda)\dagger}]$ by integrating over $\psi$ and
$\psi^{\dagger}$. The result is:

\begin{eqnarray*}
\Gamma_{eff} = - \sum_{\lambda} \eta_{\lambda} B^{(\lambda)\dagger}
B^{(\lambda)} -
{i \over 2} Tr log [{\bf 1} + 4 A^{-1} {\cal B}(A^T)^{-1} {\cal
B}^{\dagger}]
\end{eqnarray*}

\bigskip\noindent
where we have dropped some terms independent of $B$ and
$B^{\dagger}$. The mean field approximation that we shall adopt
consists in demanding that $\Gamma$ be stationary with respect to
variations of $B$ and $B^{\dagger}$:

$$
~~~~~~~~~~~~~~~~~{\delta\Gamma \over \delta B^{(\lambda)}}
 = {\delta\Gamma \over
\delta B^{(\lambda)\dagger}} = 0 ~.
$$

\bigskip
These are the gap equations whose non-trivial solutions, if any,
determine whether a condensate with the quantum numbers of
$B^{(\lambda)}$ and $B^{(\lambda)\dagger}$ can form. Because the
effective theory we are dealing with is non-renormalizable, it is
necessary to regularize these gap equations, and the size of the
gap depends, apparently, on the value of the cutoff. Thus the gap
equations in the form we derive them are not a good guide to the
size of the gap. However, they allow us to determine which channels
are attractive and hence in which channels one might expect a
condensate to form. We shall see explicitly how this works in
particular cases below.

The simplest case is only one flavor of neutrino interacting with
itself via the neutral current.  Since vector exchange produces
repulsion among like particles, we intuitively expect the $\nu-\nu$
channel to be repulsive in this case. This is borne out by our
explicit computation, which we now describe.

The starting point is the four-Fermi interaction

$$
{\cal L}_{int} = - G^2 \bar{\psi} \gamma_{\mu} \psi \bar{\psi}
\gamma^{\mu} \psi
$$

\bigskip\noindent
where $\gamma_5\psi = - \psi$. The sign is dictated by the standard
model. If we use $2$-component fermions, and make use of the
identity $\sum_{a} \sigma_{\alpha\beta}^a \sigma_{\gamma\delta}^a =
2[\epsilon_{\alpha\gamma} \epsilon_{\delta\beta} + {1 \over 2}
\delta_{\alpha\beta} \delta_{\gamma\delta}]$ we see that

\begin{eqnarray*}
{\cal L}_{int} = 2 G^2 (\psi_{\alpha}^{\dagger}
\epsilon_{\alpha\gamma} \psi_{\gamma}^{\dagger}) (\psi_{\beta}
\epsilon_{\beta\delta} \psi_{\delta}) ~.
~~~~~~~~~~~~~~~~~~~~~~~~~~~~~~~~~~~~~~~~~~~~~~~~(1)
\end{eqnarray*}

\bigskip
Making contact with our previous notation, we see that there is
only one term in the $FB$ decomposition with $Q_{\alpha\gamma} =
\sqrt{2} G \epsilon_{\alpha\gamma}$, and $\eta = - 1$. We remark that the
fact that $\eta = - 1$ already tells us that
the interaction is repulsive at the tree level; it is then to be
expected that the gap equation will not have a non-trivial
solution.

For the kinetic part of the action, again using $2$-component
notation, we have $A = i {\partial \over \partial t} - i
\vec{\sigma} \cdot \vec{\bigtriangledown} - \mu$, where $\mu$ is the
chemical potential. Hence

$$
A^{-1} = \int {d^4p \over (2\pi)^4} [{(p_0 - \mu) - \vec{p} \cdot
\vec{\sigma} \over (p_0 - \mu + i \epsilon sgn p_0)^2 - \vec{p}^2}] e^{-i p
\cdot (x - y)}
$$

\noindent
and

$$
(A^{-1})^T  = - \int {d^4p \over (2\pi)^4} {[(p_0 + \mu) - \vec{p}
\cdot
\vec{\sigma}^T] \over (p_0 + \mu + i \epsilon sgn p_0)^2 - \vec{p}^2} e^{-i
p \cdot (x - y)} ~.
$$

\bigskip\noindent
Note that the $i\epsilon$ prescription has been introduced in the
appropriate manner to take account of the role of $\mu$ as the
chemical potential. One then proceeds to construct $X \equiv 4
A^{-1} {\cal B}(A^T)^{-1} {\cal B}^{\dagger}$ which, under the
assumption (appropriate for a vacuum solution) that $B$ and
$B^{\dagger}$ are constants, can be written

$$
X = - \int {d^4p \over (2\pi)^4} {\cal F}(p) e^{-i p \cdot (x - y)}
$$

\noindent
where

$$
{\cal F}(p) = {+ 8 G^2 B^{\dagger}B \over p_0^2 - (\vec{p}
\cdot \vec{\sigma} - \mu)^2 + i \epsilon} ~.
$$

\noindent
The gap equation will involve $Tr(1 + X)^{-1}X$. Doing the $p_0$
integral in this expression provides a factor of $i$ that cancels
the explicit $i$ appearing in $\Gamma_{eff}$. The remaining
integral over $\vec{p}$ is ultraviolet divergent. We regularize
this by imposing a cutoff $\Lambda$ on the magnitude of $\vec{p}$.
This leads to the unrenormalized gap equation, written in terms of
$M = B^{\dagger}B$

$$
1 = {- G^2 \over \pi^2} \int_{- \Lambda}^{\Lambda} dp p^2 {1 \over
\sqrt{(p - \mu)^2 + 8 MG^2}} ~.
$$

\noindent
Clearly there is no solution (whereas there would have been a
solution if we had had $\eta = 1$; this would have changed the sign
of the l-h s). Note that for $M = 0$, there is a logarithmic
divergence coming from $p = \mu$. This is the singularity at the
Fermi surface mentioned above.

Including an arbitrary number $N$ of flavors turns out to be fairly
straightforward assuming no flavor mixing. One begins with the
analog of eqn. (1):

\begin{eqnarray*}
{\cal L}_{int} = 2 G^2 (\psi^{\dagger(i)}_{\alpha}
\epsilon_{\alpha\gamma} \psi_{\gamma}^{\dagger(j)}) (\psi_{\beta}^{(i)}
\epsilon_{\beta\delta} \psi_{\delta}^{(j)}) ~~
~~~~~~~~~~~~~~~~~~~~~~~~~~~~~~~~~~~~~~~~~~~~~~~~(2)
\end{eqnarray*}

\bigskip\noindent
where $i, j$ are flavor indices summed from $1$ to $N$, and
performs a separate $FB$ transformation on the flavor indices using
the identity

$$
2 \delta_{ik} \delta_{jl} = {2 \over N} \delta_{ij} \delta_{kl} +
\sum_{a = 1}^{N^2 - 1} (\lambda^a)_{ij} (\lambda^a)^*_{kl}
$$

\bigskip\noindent
where the $\lambda$'s are the generators of the fundamental
representation of $SU(N)$, normalized such that $Tr \lambda^a
\lambda^b = 2\delta^{ab}$.

Following an analysis paralleling the one-flavor case,\footnote{In
this analysis, we assume a common chemical potential for all the
flavors. What may happen if this is not the case will be discussed
briefly below.} one finds

$$
X_{\alpha\beta}^{ij} (x - y) = - M_{ij} \int {d^4p
\over (2\pi)^4} {\cal F}_{\alpha\beta}(p) e^{- ip\cdot(x - y)}
$$

\bigskip\noindent where

$$
{\cal F} = {4G^2 \over p_0^2 - (\vec{p} \cdot
\vec{\sigma} - \mu)^2 + i \epsilon}
$$

\bigskip\noindent and

$$
M_{ij} = \sum_{A,B} ~B^{(A)} ~B^{(B)\dagger}~ \lambda_{ik}^{(A)}
\lambda_{kj}^{(B)} ~.
$$

\bigskip\noindent Here the summation runs over the symmetric
$\lambda$'s, including the unit matrix suitably normalized. It is
important to note that $M$ is a positive matrix: ~$M = K
K^{\dagger}, K = \sum_{A} B^{(A)}
\lambda^{(A)}$.

This expression for $X$ leads to the following set of gap
equations:

$$
B^{(A)} = {- G^2 \over 2\pi^2} B^{(B)} \int_{- \Lambda}^{\Lambda}
dp ~p^2 tr [{1 \over \sqrt{(p - \mu)^2 + 4 G^2 M}} \lambda^{(B)}
\lambda^{(A)}] ~.
$$

\bigskip\noindent Multiply by $B^{(A)\dagger}$ and sum on $A$.
This produces

$$
\sum_{A} ~B^{(A)\dagger} ~B^{(A)} = {- G^2 \over 2 \pi^2} \int_{-
\Lambda}^{\Lambda}
dp ~p^2 tr [{1 \over \sqrt{(p - \mu)^2 + 4 G^2 M}}M]
$$

\bigskip\noindent
whose only solution is $B^{(A)} = 0$ for all $A$.

Thus, if we confine ourselves to neutrinos alone, and to the
dynamics of the standard model, we find no possibility of neutrino
pairing. Among the ways to avoid this conclusion are (a) extend the
dynamics beyond the standard model (we shall discuss this
possibility in the conclusions); (b) enlarge the dynamics to
include the charged leptons (and possibly also the quarks). We have
performed an analysis in which we consider not only neutrinos
themselves but also electrons circulating in the loop. This
generates an additional term in $\Gamma_{eff}$, and hence an
additional contribution to the gap equation for the neutrino
condensate. It does not, however, alter the result that there is no
solution to the gap equation; (c) Finally, we can consider
condensates that are composed not of neutrinos alone, but that pair
neutrinos with charged leptons. Of course, since these condensates
would be charged, their phenomenological consequences would be much
more drastic than those of purely neutrino condensates.

In any event, we begin with the Lagrange density ${\cal L} = {\cal
L}_0 + {\cal L}_{int}$, where

\begin{eqnarray*}
{\cal L}_0 &=& \bar{e} (i \bigtriangledown \!\!\!\!\!\! / - m) e +
\bar{\nu}_e i \bigtriangledown \!\!\!\!\!\! / ~~\nu_e + \bar{\nu}_{\mu} i
\bigtriangledown \!\!\!\!\!\! /
~~\nu_{\mu} - \mu_e e^{\dagger} e  \\ &-& \mu_{\nu}
\nu_e^{\dagger} \nu_e - \mu_{\nu}^{\prime}
\nu_{\mu}^{\dagger} \nu_{\mu} ~.
\end{eqnarray*}

\begin{eqnarray*}
{\cal L}_{int} &=& {- g^2 \over 8 m_W^2} [ \bar{\nu}_e \gamma_\mu
(1 - \gamma_5) e \bar{e} \gamma^{\mu} (1 - \gamma_5) \nu_e - {1
\over 2} \bar{\nu}_e \gamma_{\mu} (1 - \gamma_5) \nu_e \bar{e} \gamma^{\mu}
(1 - \gamma_5) e  \\
&+& 2 sin^2 \theta_W \bar{e}\gamma^{\mu} e \bar{\nu}_e \gamma_{\mu}
(1 - \gamma_5) \nu_e - {1 \over 2} \bar{\nu}_{\mu} \gamma_{\mu} (1
- \gamma_5) \nu_{\mu} \bar{e} \gamma^{\mu} (1 - \gamma_5) e \\
&+& 2 sin^2 \theta_W \bar{e}\gamma^{\mu} e \bar{\nu}_{\mu}
\gamma_{\mu} (1 - \gamma_5) \nu_{\mu}] ~.
\end{eqnarray*}

\bigskip
We have kept only the electron, as by far the lightest charged
lepton. We have exhibited its couplings to $\nu_e$ and $\nu_{\mu}$;
there should also be $\nu_{\tau}$, which we ignore for simplicity.
It enters in an identical manner to $\nu_{\mu}$.

Now we must implement the $FB$ transformation. It is convenient
first to make a Fierz transformation on the charged-current terms,
to get them in a form resembling the neutral current terms, and
then to make a $FB$ transformation on the result. We find

\begin{eqnarray*}
{\cal L}_{int} &=& {g^2 \over 8m_W^2} [4 sin^2 \theta_W
(\bar{\nu}_e \gamma_{\mu} \bar{e} \nu_e \gamma^{\mu} e +
\bar{\nu}_{\mu} \gamma_{\mu} \bar{e} \nu_{\mu} \gamma^{\mu} e)  \\
&+& 4(1 + 2 sin^2 \theta_W) \bar{\nu}_e \gamma^0 \gamma^2 \bar{e}
\nu_e \gamma^0 \gamma^2 e  \\
&+& 4 (- 1 + 2 sin^2 \theta_W) \bar{\nu}_{\mu} \gamma^0 \gamma^2
\bar{e} \nu_{\mu} \gamma^0 \gamma^2 e] ~.
\end{eqnarray*}

\bigskip\noindent
Note that $\gamma^0\gamma^2$ is the charge conjugation matrix, and
hence expressions of the form $\nu \gamma^0\gamma^2e$ are proper
Lorentz scalars. If we assume that the condensate will be a Lorentz
scalar (this is only for convenience: there is no particular reason
to expect that these condensates should respect Lorentz symmetry)
then we limit ourselves to the last $2$ terms. Note that they are
of opposite sign. It turns out that the term involving $\nu_e$ is
"bad" ($\eta = - 1$) whereas the one involving $\nu_{\mu}$ is
"good" ($\eta = + 1$). Hence we concentrate only on the last term,
and introduce an auxiliary field $B$ for it. We let $\kappa^2 =
{g^2 \over 2M_W^2} (1 - 2 sin^2 \theta_W)$, and for notational
convenience we introduce a doublet $\psi = (e, \nu_{\mu})$ (a
peculiar object from the point of view of the underlying standard
model), so that, for example, $m \bar{e} e = {m \over 2}
\bar{\psi} (1 +
\tau_3)
\psi$ and $\mu_e e^{\dagger}e + \mu_{\nu}^{\prime}
\nu_{\mu}^{\dagger} \nu_{\mu} = \psi^{\dagger}(\mu_1 + \mu_2 \tau_3) \psi$.
In our earlier notation, we have $A = \gamma^0 (i
\bigtriangledown \!\!\!\!\!\! /
- {m \over 2} (1 + \tau_3)) - \mu_1 - \mu_2\tau_3$ and
${\cal B} = - {\kappa^2 \over 4} B(1 - \gamma_5) \gamma^2 \gamma^0
\tau_1$.

Following the same analytic path as before, we arrive at a somewhat
more complicated gap equation:

$$
- \kappa^2  B^{\dagger} B = - {i \kappa^4 B^{\dagger} B \over \pi^3}
\int_{- \infty}^{\infty} dp ~p^2
\int_{- \infty}^{\infty} dp_0 {\cal G}(p_0,p)
$$

\bigskip\noindent where

$$
{\cal G}(p_0,p) = {p_0 - \mu_+ + p \over [(p_0 - \mu_+)^2 - p^2 -
m^2] [p_0 + \mu_- + p] - 4 \kappa^4 B^{\dagger} B [p_0 - \mu_+ +
p]} ~.
$$

\bigskip\noindent
Here $\mu_{\pm} = \mu_1 \pm \mu_2$, and the integration over poles
on the real $p_0$ axis is by the prescription $p_0 \rightarrow p_0
+ i \epsilon sgn p_0$.

It is not possible to make much analytic headway with the
expression in the general case. But if we set $\bigtriangleup^2 = 4
\kappa^4 B^{\dagger} B = 0$, the expression becomes tractable, and we
can then address two questions: ~(i) does the integral have the
correct sign to permit a solution of the gap equation? and ~(ii) is
there an infrared singularity, which would be evidence of an
instability that could be cured by setting $\bigtriangleup^2 \neq
0$? Let $\omega =
\sqrt{p^2 + m^2} > 0$. Then the $p_0$ integral can be done, yielding

\begin{eqnarray*}
1 &=& {\kappa^2 \over \pi^2} \int_{- \Lambda}^{\Lambda} dp ~p^2 \{
{\theta (\omega^2 - \mu_+^2) \over \omega} [{(\omega - p) \theta
(-p - \mu_-) \over \omega - p - 2 \mu_1} + {(\omega + p) \theta (p
+ \mu_-) \over \omega + p + 2 \mu_1}] \\ &+& {4 \mu_1 \over (\omega
+ p + 2 \mu_1)( \omega - p - 2 \mu_1)} [\theta (- \mu_+ - \omega)
\theta (- p - \mu_-) - \theta (\mu_+ + \omega) \theta (p + \mu_-)] \} ~.
\end{eqnarray*}

\bigskip\noindent
Because of the $\theta$-functions, each term in the integrand is
non-negative, thereby answering the first equation in the
affirmative.

Further, one sees that singularities can occur only at the Fermi
momentum, $p_F^2 = \mu_-^2$, and then only if the condition
$\mu_+^2 - m^2 = \mu_-^2$ is met. Since the electron density is
proportional to $[\mu_+^2 - m^2]^{3/2}$ while the neutrino density
is proportional to $\mu_-^3$ [9], this amounts to equating the
Fermi momenta of the members of the pair. Phenomenologically,
therefore, it is unlikely that this type of condensate could occur,
except maybe in the early universe when larger background densities
of both electrons and neutrinos were present.

We turn now to a consideration of the size of the condensate. A
simple $BCS$-like estimate gives

$$
\Delta \sim p_F e^{- {1 \over p_F^2 G^2}} ~.
$$

If the $G^2$ in eq. (1) or (2) is of order $G_F$, then, given the
allowed range of $p_F$, the exponential suppression makes $\Delta$
very small. One is led, therefore, to suggest the existence of a
new interaction, acting only on neutrinos, for which the effective
$G^2$ would have a scale of approximately $1$ eV instead of the
$200$ GeV characteristic of $G_F$. An interaction of the same form
as eqs. (1) or (2) would serve nicely, provided only we change the
sign, thereby generating an attractive channel. The condensate
would then produce Majorana neutrino masses of the form $m_{\nu}
\sim \Delta = G^2 \langle\nu\nu\rangle$, and possibly a
contribution to the cosmological constant $\Lambda \sim G^2 \mid
\langle \nu\nu \rangle \mid^2$. Since both $G$ and the fermi momentum $p_F$
are of order $1$ eV, one
obtains neutrino masses and a cosmological constant that are
likewise of this order.

Furthermore, if we generalize our earlier analysis and allow the
chemical potentials for the different neutrino species to vary, the
condensates could depend non-trivially on flavor, perhaps leading
to an interesting spectrum of neutrino masses and mixings.

Our conclusions can be enumerated as follows:

\bigskip\noindent
(i) There is no attractive channel in the purely neutrino sector of
the standard model;

\bigskip\noindent
(ii) The addition of charged leptons leads to attraction in the
flavor off-diagonal channels, but a pairing instability occurs only
if the Fermi momenta of the neutrino and the charged leptons are
equal;

\bigskip\noindent
(iii) In the case of neutrino-charged lepton pairing, there may
also be the possibility of condensation in a Lorentz non-invariant
channel. We have not looked at this in detail;

\bigskip\noindent
(iv) If a new interaction exists among neutrinos with
characteristic scale $1$ eV, neutrino condensates could form with
the right size to generate an interesting spectrum of masses and
mixings, as well as an appropriate contribution to the cosmological
constant. This possibility is currently under active investigation.

\vfill\eject
\noindent
{\bf Acknowledgements}

We wish to thank Fred Cooper, Gregg Gallatin, James Hormuzdiar and
Hisakazu Minakata for interesting discussions. The work of AC was
supported in part by U.S. Department of Energy Grant
\#DE-FG02-92ER-40704. The work of DGC was supported in part by U.S.
Department of Energy Grant \#DE-FG02-92ER-40741.

\bigskip\noindent
{\bf REFERENCES}

\bigskip\noindent
[1] For reviews see E. Torrente-Lujan, hep-ph/9902339, and B.
Kayser, hep-ph/9810513, and references therein.

\bigskip\noindent
[2] S. Perlmutter, et al., astro-ph/9812473 and astro-ph/9812133;
B. Schmidt, et al., Astrophys. J. {\bf 507}, 46 (1998); A.G. Riess,
et al., astro-ph/9805200; Krauss, L.M. and Turner, M.S., Gen. Rel.
Grav. {\bf 27}, 1137 (1995); Ostriker, J.P. and Steinhardt, P.,
Nature {\bf 377}, 600 (1995).

\bigskip\noindent
[3] M. Alford, K. Rajagopal and F. Wilczek, Phys. Lett. {\bf B422},
247 (1998); R. Rapp, T. Shafer, E.V. Shuryak and M. Velkovski,
Phys. Rev. Lett. {\bf 81}, 53 (1998); T. Schafer, Nucl. Phys. {\bf
A462}, 45 (1998); J. Berges and K. Rajagopal, Nucl. Phys. {\bf
B538}, 215 (1999).

\bigskip\noindent
[4] M. Alford, K. Rajagopal and F. Wilczek, Nucl. Phys. {\bf B537},
443 (1999); R. Pisarski and D. Rischke, nucl-th/9811104; D.T. Son,
hep-ph/9812287.

\bigskip\noindent
[5] A. Chodos, H. Minakata and F. Cooper, Phys. Lett. B (to be
published), hep-ph/9812305; and manuscript in preparation.

\bigskip\noindent
[6] P.B. Pal and K. Kar, hep-ph/9809410.

\bigskip\noindent
[7] V.L. Ginzburg and G.F. Zharkov, ZhETF Pi\'{s}ma {\bf 5}, 275
(1967) [JETP Letters {\bf 5}, 223 (1967)]; see also V. Alonso, J.
Chela-Flores and R. Paredes, Nuov. Cim. {\bf 67B}, 213 (1982).

\bigskip\noindent
[8] N. Evans, S.D.H. Hsu and M. Schwetz, hep-ph/9808444 and
hep-ph/9810515; T. Shafer and F. Wilczek, hep-ph/9810509; D.T. Son
(ref. 4); R. Shankar, Rev. Mod. Phys. {\bf 66}, 129 (1993); J.
Polchinski, TASI Lectures (1992) (hep-th/9210046).

\bigskip\noindent
[9] See for example, A. Chodos, K. Everding and D.A. Owen, Phys.
Rev. {\bf D42}, 2881 (1990), sec 2.

\end{document}